\documentclass[11pt]{article}

\def\slashchar#1{\setbox0=\hbox{$#1$}
   \dimen0=\wd0 \setbox1=\hbox{/} \dimen1=\wd1
   \ifdim\dimen0>\dimen1 \rlap{\hbox to \dimen0{\hfil/\hfil}} #1
   \else  \rlap{\hbox to \dimen1{\hfil$#1$\hfil}} / \fi}

\usepackage{graphicx}
\usepackage{cite}

\begin{document}

\title{\bf Large-$N_c$ Regge models and the $\langle A^2 \rangle$ condensate}

\author{{\bf Wojciech Broniowski$^{1,2}$ and Enrique Ruiz Arriola$^{3}$} \\~\\
$^1$Institute of Nuclear Physics PAN, PL-31342~Cracow, Poland\\
$^2$Institute of Physics, \'Swi\c{e}tokrzyska Academy,  PL-25406~Kielce, Kielce, Poland\\ 
$^3$Departamento de F\'{\i}sica At\'omica, Molecular y Nuclear, Universidad de Granada \\E-18071 Granada, Spain}


\date{Talk presented by WB at the Mini-Workshop Bled 2007: \\HADRON STRUCTURE AND LATTICE QCD, Bled (Slovenia), 9-16 July 2007}

\maketitle

\vspace{-2mm}

\abstract{We explore the role of the $\langle A^2 \rangle$ gluon
condensate in matching Regge models to the operator product expansion
of meson correlators.}

\vspace{5mm}

This talk is based on Ref.~\cite{RuizArriola:2006gq}, where the
details may be found. The idea of implementing the principle of
parton-hadron duality in Regge models has been discussed in
Refs.~\cite{Golterman:2001nk,Beane:2001uj,%
Simonov:2001di,Golterman:2002mi,Afonin:2003gp,Afonin:2004yb,Afonin:2006sa}.
Here we carry out this analysis with the dimension-2 gluon condensate
present. The dimension-two gluon condensate, $\langle A^2 \rangle$,
was originally proposed by Celenza and Shakin~\cite{Celenza:1986th}
more than twenty years ago. Chetyrkin, Narison and
Zakharov~\cite{Chetyrkin:1998yr} pointed out its sound
phenomenological as well as theoretical
\cite{Gubarev:2000eu,Gubarev:2000nz,Kondo:2001nq,Verschelde:2001ia,Capri:2006ne} consequences. Its
value can be estimated by matching to results of lattice calculations
in the Landau gauge \cite{Boucaud:2001st,RuizArriola:2004en}, and
their significance for non-perturbative signatures above the
deconfinement phase transition was analyzed in~\cite{Megias:2005ve}.
Chiral quark-model calculations were made in~\cite{Dorokhov:2003kf}
where $\langle A^2 \rangle$ seems related to constituent quark
masses. In spite of all this flagrant need for these unconventional
condensates the dynamical origin of $\langle A^2 \rangle$ remains
still somewhat unclear; for recent reviews see, {\em e.g.},
\cite{Zakharov:2005cg,Narison:2005hb}.

For large $Q^2$ and fixed $N_c$ the modified OPE (with the $1/Q^2$
term present) for the chiral combinations of the transverse parts of
the vector and axial currents is
\begin{eqnarray}
&&\Pi^T_{V+A} (Q^2) = \frac{1}{4\pi^2} \Big\{-\frac{N_c}{3}
\log \frac{Q^2}{\mu^2} {-\frac{\alpha_S}{\pi}\frac{\lambda^2}{Q^2}} + \frac{\pi}{3}
\frac{\langle \alpha_S G^2 \rangle}{Q^4} + \dots \Big\} \nonumber \\
&&\Pi^T_{V-A} (Q^2) = - \frac{32 \pi }{9}\frac{\alpha_S \langle \bar q q\rangle^2}{Q^6}+\dots \label{qcd}
\end{eqnarray} 
On the other hand, at large-$N_c$ and any $Q^2$ these correlators may
be saturated by infinitely many mesonic states,
\begin{eqnarray}
\Pi^T_V(Q^2) \!= \!\sum_{n=0}^\infty \!\frac{F_{V,n}^2}{M_{V,n}^2+ Q^2} + c.t., \;\; \Pi^T_A(Q^2) \!= \!\frac{f^2}{Q^2} + 
\sum_{n=0}^\infty\! \frac{F_{A,n}^2}{M_{A,n}^2+ Q^2}+c.t. \label{meson}
\end{eqnarray} 
The basic idea of parton-hadron duality is to match Eq.~(\ref{qcd})
and (\ref{meson}) for both large $Q^2$ and $N_c$ (assuming that both
limits commute).  We use the radial Regge spectra, which are well
supported experimentally \cite{Anisovich:2000kx}
\begin{eqnarray} 
M^2_{V,n} = M_{V}^2 + a_V n, \;\;\; M^2_{A,n} = M_{A}^2 + a_A n, \;\;\; n=0,1,\dots \label{regge} 
\end{eqnarray}
The vector part, $\Pi^T_V$, satisfies the once-subtracted dispersion relation
\begin{eqnarray}
\hspace{-4mm}\Pi^T_V(Q^2) &=& \sum_{n=0}^\infty \left ( \frac{F_{V,n}^2}{M_{V}^2 +
a_V n + Q^2} -  \frac{F_{V,n}^2}{M_{V}^2+a_V n} \right ).
\end{eqnarray} 
We need to reproduce the $\log Q^2$ in OPE, for which only the
asymptotic part of the meson spectrum matters. This leads to the
condition that at large $n$ the residues become independent of $n$,
$F_{V,n} \simeq F_V$ and $F_{A,n} \simeq F_A$.  Thus all the
highly-excited radial states are coupled to the current with equal
strength!  Or: asymptotic dependence of $F_{V,n}$ or $F_{A,n}$ on $n$
would damage OPE. Next, we carry out the sum explicitly (the dilog
function is $\psi(z)=\Gamma'(z)/\Gamma(z)$)
\begin{eqnarray} 
&& \sum_{n=0}^\infty \left ( \frac{F_i^2}{M_i^2 +
a_i n + Q^2} -  \frac{F_i^2}{M_i^2+a_i n} \right ) =
\frac{F_i^2}{a_i} \left [ \psi\left(\frac{M_i^2}{a_i}\right)\!-\!\psi\left(\frac{M_i^2+Q^2}{a_i}\right) \right ] \nonumber \\
&& = \frac{F_i^2}{a_i} \left [\!-\!\log\left(\frac{Q^2}{a_i}\right)\!+\!\psi\left(\frac{M_i^2}{a_i}\right)\!+\!
\frac{a_i-2 M_i^2}{2 Q^2}+\frac{6 M_i^4-6 a_i M_i^2+a_i^2}{12 Q^4} \!+\!\dots \right ],
\end{eqnarray}
where $i=V,A$.  $\Pi_{V-A}$ satisfies the unsubtracted dispersion
relation (no $\log Q^2$ term), hence
\begin{eqnarray} 
{F_V^2}/{a_V}={F_A^2}/{a_A}. 
\end{eqnarray}
This complies to the chiral symmetry restoration in the high-lying
spectra \cite{Glozman:2002cp,Glozman:2003bt}.  Further, we assume
$a_V=a_A=a$, or $F_V=F_A=F$, which is well-founded experimentally, as
$\sqrt{\sigma_A} = 464 {\rm MeV}$, $\sqrt{\sigma_V} = 470 {\rm MeV}$
\cite{Anisovich:2000kx}.

The simplest model we consider has strictly linear trajectories all
the way down,
\begin{eqnarray}
&&\hspace{-5mm} \Pi^T_{V-A}(Q^2) = \frac{F^2}{a} 
\left [ - \psi \left ( \frac{M_{V}^2+Q^2}{a}
\right ) +
\psi \left ( \frac{M_{A}^2+Q^2}{a} \right ) \right ] - \frac{f^2}{Q^2}\,
\nonumber \\ &&
=  \left ( \frac{F^2}{a} (M_{A}^2-M_V^2)-f^2 \right ) 
\frac{1}{Q^2}  +
\left ( \frac{F^2}{2a} (M_{A}^2-M_V^2)(a-M_A^2-M_V^2) \right ) 
\frac{1}{Q^4}+ \dots \nonumber
\end{eqnarray}
Matching to OPE yields the two Weinberg sum rules:
\begin{eqnarray}
 f^2 &=& \frac{F^2}{a} (M_{A}^2-M_V^2), \hspace{27mm} {\rm (WSR~I)} \nonumber \\
 0&=&(M_{A}^2-M_V^2)(a-M_A^2-M_V^2). \hspace{7mm} {\rm (WSR~II)} \nonumber  
\end{eqnarray}
The $V+A$ channel needs regularization. We proceed as follows: carry $d/dQ^2$, compute the convergent sum, and integrate
back over $Q^2$. The result is
\begin{eqnarray}
&&\hspace{-5mm}\Pi^T_{V+A}(Q^2) = \frac{F^2}{a} \left [ - \psi \left ( \frac{M_{V}^2+Q^2}{a}
\right ) -\psi \left ( \frac{M_{A}^2+Q^2}{a} \right ) \right ] + \frac{f^2}{Q^2} + {\rm const.}  = -\frac{2F^2}{a} \log \frac{Q^2}{\mu^2} \nonumber \\ &&
+ \left ( f^2 +F^2 -\frac{F^2}{a} (M_A^2+M_V^2)   \right ) \frac{1}{Q^2} 
 +
\frac{F^2}{6a}\left (a^2 -3 a (M_A^2+M_V^2)+3(M_A^4+M_V^4)  \right ) 
\frac{1}{Q^4}+ \dots \nonumber
\end{eqnarray}
Matching of the coefficient of $\log Q^2$ 
to OPE gives the relation
\begin{eqnarray} 
a= 2\pi \sigma = \frac{{24\pi^2} F^2}{N_c},
\end{eqnarray}
where $\sigma$ denotes the (long-distance) string tension. From the $\rho
\to 2 \pi$ decay one extracts \mbox{$F = 154~{\rm MeV}$} \cite{Ecker:1988te}
which gives $\sqrt{\sigma} = 546~{\rm MeV}$, compatible to the value
obtained in lattice simulations: $\sqrt{\sigma} = 420~{\rm MeV}$
\cite{Kaczmarek:2005ui}. Moreover, from the Weinberg sum rules
\begin{eqnarray} 
M_A^2=M_V^2+\frac{24\pi^2}{N_c} f^2, \;\;\; a=M_A^2+M_V^2=2M_V^2+\frac{24\pi^2}{N_c} f^2.
\end{eqnarray}
Matching higher twists fixes the dimension-2 and 4 gluon condensates:
\begin{eqnarray}
-\frac{\alpha_S \lambda^2}{4\pi^3}=f^2, \;\;\;\; \frac{\alpha_S \langle G^2 \rangle}{12\pi}=
\frac{M_A^4-4M_V^2 M_A^2 +M_V^4}{48 \pi^2}. 
\end{eqnarray}
Numerically, it gives $-\frac{\alpha_S \lambda^2}{\pi}=0.3~{\rm
GeV}^2$ as compared to $0.12 {\rm GeV}^2$ from
Ref.~\cite{Chetyrkin:1998yr,Zakharov:2005cg}.  The short-distance
string tension is $\sigma_0 = - 2 \alpha_s \lambda^2/N_c = 782~{\rm
MeV}$, which is twice as much as $\sigma$. The major problem of the
strictly linear model is that the dimension-4 gluon condensate is
negative for $M_V \ge 0.46$~GeV. Actually, it never reaches the QCD
sum-rules value. Thus, the strictly linear radial Regge model is {\em
too restrictive}!

We therefore consider a modified Regge model where for low-lying states both
their residues and positions may depart from the linear trajectories. 
The OPE condensates are expressed in terms of the
parameters of the spectra.  A very simple modification moves only the
position of the lowest vector state, the $\rho$ meson.
\begin{eqnarray} 
&&{M_{V,0}=m_\rho,} \;\;\; M_{V,n}^2=M_V^2+a n, \;\;\;\; n \ge 1 \nonumber \\
&& M_{A,n}^2=M_A^2+a n, \;\;\;\; n \ge 0.
\end{eqnarray}
For the Weinberg sum rules (we use $N_c=3$ from now on)
\begin{eqnarray} 
M_A^2=M_V^2+8\pi^2 f^2, \;\;\;\;\; a = 8\pi^2 F^2=\frac{8\pi^2 f^2 \left(4 \pi ^2 f^2+{M_V}^2\right)}{4 \pi ^2 f^2-{m_\rho}^2+{M_V}^2}.
\end{eqnarray}
We fix $m_\rho=0.77~{\rm GeV}$, and  $\sigma$ is the only free parameter of the model. Then
\begin{eqnarray} 
&&M_V^2=\frac{-16 \pi ^3 f^4+4 \pi ^2 \sigma f^2-{m_\rho}^2 \sigma}{4 f^2 \pi - \sigma}, \;\; -\frac{\alpha_S \lambda^2}{4\pi^3}=\frac{16 \pi ^3 f^4-\pi  \sigma^2+{m_\rho}^2 \sigma}{16 f^2 \pi ^3-4 \pi ^2 \sigma}, \nonumber \\
&& \frac{\alpha_S \langle G^2 \rangle}{12\pi}=2 \pi ^2 f^4-\pi \sigma f^2 + \frac{3 \sigma \left(\frac{{m_\rho}^2
   \sigma}{\left(\sigma-4 f^2 \pi \right)^2}-2 \pi \right) {m_\rho}^2}{8\pi ^2}+\frac{\sigma^2}{12}. 
\end{eqnarray}
\begin{figure}[tb]
\vspace{-13mm}
\begin{center}
\includegraphics[angle=0,width=0.4\textwidth]{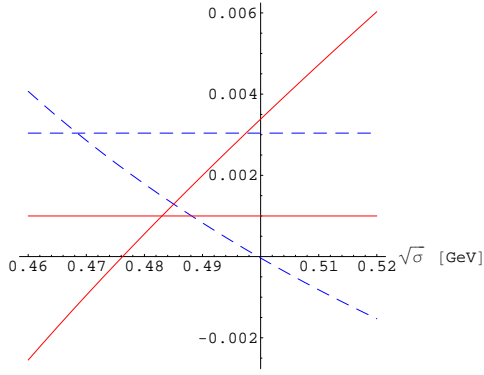}
\end{center}
\vspace{-11mm}
\caption{\footnotesize Dimension-2 (solid line, in GeV$^2$) and -4 (dashed line,
in GeV$^4$) gluon condensates plotted 
as functions of the square root of the string tension. The straight lines 
indicate phenomenological estimates. The fiducial region in $\sqrt{\sigma}$ 
for which both condensates are positive is in the acceptable range compared to 
the values of Ref.~\cite{Anisovich:2000kx} and other studies.}
\label{fig:gg}
\end{figure}
The window for which both condensates are positive yields very
acceptable values of $\sigma$.  The consistency check of near equality
of the long- and short-distance string tensions, $\sigma \simeq
\sigma_0$, holds for $\sqrt{\sigma} \simeq 500 {\rm MeV}$.  The
magnitude of the condensates is in the ball park of the ``physical''
values.  The value of $M_V$ in the ``fiducial'' range is around
$820$~MeV.  The experimental spectrum in the $\rho$ channel is has
states at 770, 1450, 1700, 1900$^\ast$, and 2150$^\ast$~MeV, while the
model gives 770, 1355, 1795, 2147~MeV (for $\sigma=(0.47~{\rm
GeV}^2$).  In the $a_1$ channel the experimental states are at 1260
and 1640~MeV, whereas the model yields 1015 and 1555~MeV.

We note that the $V-A$ channel well reproduced with radial Regge
models.  The Das-Mathur-Okubo sum rule gives the Gasser-Leutwyler
constant $L_{10}$, while the Das-Guralnik-Mathur-Low-Yuong sum rule
yields the pion electromagnetic mass splitting.  In the strictly
linear model with $M_A^2 \!=\! 2 M_V^2$ and $M_V\!=\!\sqrt{24
\pi^2/N_c} f \!=\!764~{\rm MeV}$ we have $\sqrt{\sigma}=
\sqrt{3/2\pi} M_V = 532~{\rm MeV}$, $F= \sqrt{3} f = 150~{\rm MeV}$, $
L_{10}=-N_c/( 96 \sqrt{3} \pi)= -5.74 \times 10^{-3} {( -5.5 \pm 0.7
\times 10^{-3})_{\rm exp}}$, $ m^2_{\pi^\pm}-m^2_{\pi_0}= (31.4~{\rm
MeV})^2 \; {(35.5~{\rm MeV})^2_{\rm exp}}$.  In our second model with
$\sigma=(0.48~{\rm GeV})^2$ we find $L_{10}=-5.2 \times 10^{-3} $ and
$ m^2_{\pi^\pm}-m^2_{\pi_0} = (34.4~{\rm MeV})^2$.

To conclude, let us summarize our results and list some further
related studies.

\begin{itemize}

\item Matching OPE to the radial Regge models produces in a natural
way the $1/Q^2$ correction to the $V$ and $A$ correlators.
Appropriate conditions are satisfied by the asymptotic spectra, while
the parameters of the low-lying states are tuned to reproduce the
values of the condensates.

\item In principle, these parameters of the spectra are measurable, 
hence the information encoded in the low-lying states is the same as the
information in the condensates.

\item Yet, sensitivity of the values of the condensates to the 
parameters of the spectra, as seen by comparing the two
explicit models considered in this paper, makes such a study difficult 
or impossible at a more precise level.

\item Regge models work very well in the $V-A$
channel. In~\cite{Arriola:2006sv} it is shown how the spectral (in
fact chiral) asymmetry between vector and axial channel is generated
via the use of $\zeta$-function regularization for {\it each} channel
separately.

\item We comment that effective low-energy chiral models produce 
$1/Q^2$ corrections ({\em i.e.} provide a scale of dimension 2), 
{\em e.g.}, the instanton-based chiral quark model gives \cite{Dorokhov:2003kf}  
\begin{eqnarray} 
-\frac{\alpha_S}{\pi}\lambda^2  
=-2N_{c}\int du\frac{u}{u+M(u)^2 }M\left( u\right) M^{\prime }\left(
u\right) \simeq 0.2{\rm ~GeV}^{2}.
\end{eqnarray}

\item In the presented Regge approach the pion
distribution amplitude is constant, $\phi(x)=1$, at the low-energy
hadronic scale, similarly as in chiral quark models
\cite{RuizArriola:2006ii}.

\end{itemize}


\end{document}